\documentclass[11pt]{article}
\usepackage{amsmath}
\usepackage{amssymb}
\advance\hoffset by -7mm   
\makeatletter 
\@addtoreset{equation}{section}  
\makeatother 
 
\def\ii{\'\i}

\def\ftoday{{\sl {Le \number\day \space\ifcase\month 
\or janvier\or f\'evrier\or mars\or avril\or mai
\or juin\or juillet\or ao\^ut\or septembre\or octobre
\or novembre \or d\'ecembre\fi\space \number\year}}}    
\def\ptoday{{\sl {\number\day \space de\space \ifcase\month 
\or janeiro\or fevereiro\or mar{\c c}o\or abril\or maio
\or junho\or julho\or agosto\or setembro\or outubro
\or novembro \or dezembro\fi\space de\space \number\year}}}    
\def\gtoday{{\sl {Den \number\day. \ifcase\month 
\or Januar\or Februar\or M\"arz\or April\or Mai
\or Juni\or Juli\or August\or September\or Oktober
\or November \or Dezember\fi\space \number\year}}}    
\def\today{{\sl {\ifcase\month
\or January\or February\or March\or April\or May
\or June\or July\or August\or September\or October
\or November \or December\fi \space\number\day,\space 
                                            \number\year}}}
\newcommand{\journal}[4]{{\em #1~}#2\,(#3)\,#4}

\newcommand{\ijmp}{\journal {Int. J. Mod. Phys.}}

\newcommand{\pr}{\journal {Phys. Rev.}}

\newcommand{\cmp}{\journal {Commun. Math. Phys.}}

\newcommand{\cqg}{\journal {Class. Quantum Grav.}}

\newcommand{\np}{\journal {Nucl. Phys.}}

\newcommand{\annp}{\journal {Ann. Phys. (N.Y.)}}

\newcommand{\GRG}{\journal {Gen. Rel. Grav}}
\renewcommand{\a}{\alpha}
\renewcommand{\b}{\beta}
\newcommand{\g}{\gamma}           \newcommand{\GA}{\Gamma}
\renewcommand{\d}{\delta}         
\newcommand{\e}{\varepsilon}
\newcommand{\ka}{\kappa}
\newcommand{\la}{\lambda}        \newcommand{\LA}{\Lambda}
\newcommand{\m}{\mu}
\newcommand{\n}{\nu}
\newcommand{\om}{\omega}         
\newcommand{\s}{\sigma}           \renewcommand{\S}{\Sigma}
\renewcommand{\AA}{{\cal A}}

\newcommand{\FF}{{\cal F}}
\newcommand{\GG}{{\cal G}}
\newcommand{\HH}{{\cal H}}

\newcommand{\MM}{{\cal M}}

\newcommand{\PP}{{\cal P}}

\newcommand{\TT}{{\cal T}}

\newcommand{\XX}{{\cal X}}
\newcommand{\YY}{{\cal Y}}

\newcommand{\esp}{\\[3mm]}

\newcommand{\sla}{\raise.15ex\hbox{$/$}\kern -.57em} 
\newcommand{\Sla}{\raise.15ex\hbox{$/$}\kern -.70em}

\def\LP{\displaystyle{\Biggl(}}
\def\RP{\displaystyle{\Biggr)}}
\newcommand{\lp}{\left(}\newcommand{\rp}{\right)}

\newcommand{\complex}{{\kern .1em {\raise .47ex
\hbox {$\scriptscriptstyle |$}}
    \kern -.4em {\rm C}}}
\newcommand{\real}{{{\rm I} \kern -.19em {\rm R}}}
\newcommand{\rational}{{\kern .1em {\raise .47ex
\hbox{$\scripscriptstyle |$}}
    \kern -.35em {\rm Q}}}
\renewcommand{\natural}{{\vrule height 1.6ex width
.05em depth 0ex \kern -.35em {\rm N}}}

\newcommand{\trace}{{\rm {Tr} \,}}
\newcommand{\half}{\frac{1}{2}}
\newcommand{\pa}{\partial}

\newcommand{\dfud}[2]{{\displaystyle{\frac{\delta #1}{\delta #2}}}}
\renewcommand{\dfrac}[2]{{\displaystyle{\frac{#1}{#2}}}}
\newcommand{\dsum}[2]{\displaystyle{\sum_{#1}^{#2}}}   
\newcommand{\dint}{{\displaystyle{\int}}}

\newcommand{\ie}{{{\em i.e.},\ }}

\newcommand{\twiddle}{\lower.9ex\rlap{$\kern -.1em\scriptstyle\sim$}}
\newcommand{\vev}[1]{\langle {#1}\rangle}

\newcommand{\equ}[1]{(\ref{#1})}
\newcommand{\eq}{\begin{equation}}
\newcommand{\eqn}[1]{\label{#1}\end{equation}}
\newcommand{\eea}{\end{eqnarray}}
\newcommand{\eqa}{\begin{eqnarray}}
\newcommand{\eqan}[1]{\label{#1}\end{eqnarray}}
\newcommand{\ba}{\begin{array}}
\newcommand{\ea}{\end{array}}
\newcommand{\eqac}{\begin{equation}\begin{array}{rcl}}
\newcommand{\eqacn}[1]{\end{array}\label{#1}\end{equation}}

\newcommand{\bz}{\begin{enumerate}}
\newcommand{\ez}{\end{enumerate}}

\newcommand{\bsJ}{{\boldsymbol{J}}}
\newcommand{\bsP}{{\boldsymbol{P}}}
\newcommand{\bsQ}{{\boldsymbol{Q}}}
\newcommand{\bsR}{{\boldsymbol{R}}}



\newcommand{\SADS}{S(A)dS}
\newcommand{\ADS}{(A)dS}

\newcommand{\sads}{\textsf{s(a)ds}}
\newcommand{\ads}{\textsf{(a)ds}}
\newcommand{\ds}{\textsf{ds}}
\newcommand{\bfA}{\mathbf{A}}
\newcommand{\bfB}{\mathbf{B}}
\newcommand{\bfF}{\mathbf{F}}
\newcommand{\bfD}{\mathbf{D}}
\newcommand{\bfd}{\mathbf{d}}
\newcommand{\bfx}{\mathbf{x}}
\newcommand{\bfAA}{\mathbf{\mathcal{A}}}

\begin{document}

%%%%%%%%%%%%%%%%%%%%%%%%%%%%%%%%%%%%%%%%%%%%%%%5555
\title{Loop Quantization of a Model for $D=1+2$ (Anti)de Sitter  Gravity  Coupled  to Topological Matter}

\author{Clisthenis P. Constantinidis \\
{\small Departamento de F\ii sica, Universidade Federal do Esp\'{\i}rito Santo (UFES)}\\
{\small Vit\'oria, ES, Brazil.}\esp
Zui Oporto and Olivier Piguet\\
{\small Departamento de F\ii sica, 
Universidade Federal de Vi\c cosa -- UFV, Vi\c cosa, MG, Brazil}
}

%\date{\today}
\date{April 15, 2014}

\maketitle

\begin{center} 

\vspace{-5mm}

{\small\tt 
E-mails: cpconstantinidis@pq.cnpq.br, 
azurnasirpal@gmail.com,
opiguet@pq.cnpq.br}
\end{center}

%%%%%%%%%%%%%%%%%%%%%%%%%%%%%%%%%%%%%%%%%%%%%%%%%%%%%%%%%%%%%%%%%%
\vspace{5mm}

\begin{abstract}
We present a complete quantization of Lorentzian $D=1+2$ gravity with cosmological constant, coupled to a set of topological matter fields. The approach of Loop Quantum Gravity is used thanks to a partial gauge fixing leaving a residual gauge invariance under a {\it compact} semi-simple gauge group,
namely Spin(4) = SU(2) $\times$ SU(2). A pair of   quantum  observables is constructed, which are non-trivial despite of being null at the classical level.
\end{abstract}

%%%%%%%%%%%%%%%%%%%%%%%%%%%%%%%%%%%%%%%%%%%%%%%%%%%%%%%%%%%%%%%%%%
\section{Introduction}

This paper presents a generalization of a previous work~\cite{BCOP} where the Loop Quantum Gravity (LQG) quantization of $D=1+2$ gravity with a positive cosmological constant, in the presence of a
Barbero Immirzi-like parameter 
analogous to the one which may be introduced in the four dimensional gravitation 
theory~\cite{barbero-immirzi,holst} (and first introduced in the 
three-dimensional theory by the authors 
of~\cite{bonzom-livine}) 
was performed using a partial gauge fixing procedure leaving the compact SU(2) group as the residual group of gauge invariance. 

$D=2+1$ gravity with a cosmological constant $\LA$ is described by 
 a Lorentz connection $\om$ and a triad $e$ 1-forms, components of an \ads\ 
 connection~\cite{Witten:1988hc}. \ADS\ denotes the $D=1+2$ de Sitter dS = SO(1,3) or anti-de Sitter 
 ADS = SO(2,2) group, and \ads,  \ds\ = so(1,3) or {\textsf{ads}} = so(2,2), its Lie algebra. 
  The canonical structure and quantization of this theory have been studied, beyond the pioneering work of Witten~\cite{Witten:1988hc}, 
in~\cite{bonzom-livine,Geiller1,Geiller2,Geiller3,Buffenoir}, 
among others (see~\cite{Carlip} for a general review based on previous literature). 
A Barbero Immirzi-like parameter has also been defined in~\cite{Geiller1}
for the three-dimensional theory, although in a different way as in~\cite{bonzom-livine}, and its role has been discussed 
in~\cite{Geiller2,Geiller3}  as well for the classical as for the 
quantum theory\footnote{We thank Marc Geiller for informing us 
on the references~\cite{Geiller1,Geiller2,Geiller3}.}. 
 
 The coupling to ``topological matter'' shown in the present paper  
will be   per\-for\-med via an extension  of the \ads\ Lie algebra   which consists in  the addition of a multiplet  of 
 {\it non-commuting} generators belonging to the adjoint representation of \ads, in such a way that the resulting algebra closes on a semi-simple algebra, denoted by \sads\ for ``semi-simple extension of \ads''. It results that this extension is a deformation of an algebra introduced by 
the authors 
of~\cite{carlip-gegenberg,carlip-gegenberg-mann,freidel-mann-popescu} as the extension of \ads\ by {\it commuting} generators in the adjoint representation. The deformation parameter, $\la$, will play the role of a coupling constant.  This extended algebra possesses four 
non-degenerate invariant quadratic forms, instead of 2 for \ads, which 
will imply the presence of  four independent couplings, three of them
 being generalized Barbero-Immirzi like parameters.
 
The theory   will be   defined as the Chern-Simons theory of a \sads\ connection, the components of which are the gravity fields:  the spin connection $\om$ and the triad $e$; and a multiplet of matter fields: 1-forms $\{b,\,c\}$ transforming in the adjoint representation of \ads. For suitable choices of the signs of the \sads\ structure constant parameters  $\LA$ and $\la$, the algebra admits so(4) as a compact sub-algebra. 
We shall restrict ourselves to this family of parametrizations.
 Moreover,   with the same choice of signs,  it 
factorizes as the direct sum of two \ds\ sub-algebras, which allows a simpler 
treatment of the theory, and in particular permits us to use the results 
of~\cite{BCOP} where the pure gravity case, based on the \ADS\ gauge group, is studied in details.
 
  Loop quantization methods will be applied to the canonical quantization of the model, in the special case of the 2-dimensional space sheet topology being that of a cylinder.  A partial gauge fixing preserving gauge invariance under Spin(4), the universal covering of SO(4),   will have  to be performed. The constraints   will be  curvature constraints which can be entirely solved, leaving a physical Hilbert space, with a spin network type basis labelled by pairs of half-integer spins.
Finally a pair of   quantum  ob\-ser\-vab\-les will be constructed, which are diagonal in the 
spin-network basis, with a discrete spectrum reminiscent of the area operator spectrum of dimension $1+3$ 
Loop Quantum Gravity (LQG)~\cite{general-ref}.
 
 The model is presented in Section~\ref{model} in the canonical formalism,   with the derivation of the Hamiltonian  and of the constraints.  In Section~\ref{calibre-axial} 
we gauge fix the non-compact part of the gauge group, leaving an so(4) residual gauge invariance, which allows us, in Section \ref{quantization}, to proceed to the quantization using the standard tools of LQG. 
Observables are constructed in Section~\ref{observavel}. The Appendix is 
devoted to the definition of the semi-simple extension of a Lie algebra,
with application to the extension \sads\ of the \ads\ algebra together with  the study of its compact sub-algebras and factorization properties.

%%%%%%%%%%%%%%%%%%%%%%%%%%%%%%%%%%%%%%%%%%%%%%%%%%%%%%%%%%%%%%%%%%%
\section{A model for (anti-)de Sitter gravity with topological matter}
\label{model}

The model is described as a Chern-Simons theory 
in a  $D=1+2$ orientable 
manifold  $\mathcal{M}$. The gauge group is the "semi-simple extension" \SADS\ of the  $D=1+2$ de Sitter or anti-de Sitter group \ADS\ = SO(1,3) or SO(2,2)  with the corresponding Lie algebra \sads\  being  described in the Appendix. We consider as a basis the six generators $\{J^I,P^I;I=0,1,2\}$ of \ADS, together with the six extension generators $\{Q^I,R^I;I=0,1,2\}$, satisfying the commutation rules
\begin{equation}\begin{array}{llll}
[J^I,J^J]= \e^{IJ}{}_K J^K\,,\quad &[J^I,P^J]= \e^{IJ}{}_K P^K\,,
\quad&[P^I,P^J]= \s\LA\,\e^{IJ}{}_K J^K\,,& \esp
[J^I,Q^J]= \e^{IJ}{}_K Q^K\,,\quad& [J^I,R^J]= \e^{IJ}{}_K R^K\,,&&\esp
[P^I,Q^J]= \e^{IJ}{}_K R^K\,,\quad&[P^I,R^J]
= \s\LA\,\e^{IJ}{}_K Q^K\,,&&\esp
[Q^I,Q^J]=\s \la\, \e^{IJ}{}_K J^K\,,\quad&[Q^I,R^J]= \s {\la}\,\e^{IJ}{}_K P^K\,,
\quad&[R^I,R^J]= \LA \la\,\e^{IJ}{}_K J^K\,.&
\end{array}
\label{SADS-algebra}\end{equation}
$\s$ is the $D=1+2$ metric signature\footnote{The indices $I,J,\dots$ take 
the values 0,1,2. They may be lowered or raised with the metric 
$\eta_{IJ}={\rm diag}(\s,1,1)$, $\s=\pm1$ being the signature of 
the rotation or Lorentz group SO(3) or SO(1,2). 
The completely antisymmetric tensor $\e^{IJK}$ is defined by $\e^{012}=1$. 
Note that 
$\e_{012}=\eta_{0I}\eta_{1J}\eta_{2K} \e^{IJK} =\s$.\\
Space-time indices will be denoted later on by greek letters 
$\m,\n,\dots = 0,1,2$ or the symbols $t,x,y$, and space indices 
by latin letters $a,b,\dots = 1,2$ or the symbols $x,y$.}, 
$\LA$ and ${\la}$ are two arbitrary parameters defining the 
closure of the algebra, which will play in turn the   roles of 
a cosmological constant and of a coupling constant,  as we shall see. The properties of this 
algebra are described in the 
Appendix.
  
\noindent\textbf{Remark.}
The present model is a generalization of the model of 
Refs.~\cite{carlip-gegenberg,carlip-gegenberg-mann,freidel-mann-popescu} in the sense that, for $\LA={\la}=0$, the algebra \equ{SADS-algebra} reduces to the Lie algebra of the  gauge group I(ISO(1,2)) -- an extension of the Poincar\'e group ISO(1,2)  through Abelian 
generators\footnote{We use the notation of~\cite{freidel-mann-popescu} for the basis generators.}.

The field content of the theory is given by the \sads\ connection 1-form
\eq 
\AA = \om^I J_I + e^I P_I + b^I Q_I + c^I R_I
\equiv \dsum{a=1}{12}\AA^\a \TT_\a  \,.
\eqn{sads-connection}
In order to write an action, we need an \sads-invariant 
  non-degenerate  quadratic form.
It turns out that in the present case we have 4 such forms,    $K^{i}_{\a\b}$,  (given in 
Eqs. \equ{4-quad-forms} of the Appendix) and then the action may be written as 
%.... 
%
%( TIRARIA ISSO Thus we   must  write the action as )  
%
%...
a superposition of four Chern-Simons 
actions for the connection \equ{sads-connection}, each one corresponding 
to one   of these  quadratic forms\footnote{We don't write explicitly the wedge symbol
$\wedge$ for the external products of forms.}: 
\eq 
S= \dsum{i=1}{4} c_i S_i\,,\quad
S_i= \dint_\MM K^{i}_{\a\b}\, \AA^\a  \lp d\AA + \frac23\AA\AA \rp^\b\
\eqn{gen-action-JPQR}
%\nonindent \textbf{Remark:} 
It is interesting to explicitly write the second term:
\[
S_2 = \dint_\MM \lp e^I F_I(\om) + \dfrac{\s\LA}{6} e^I(e\times e)_I
+\s {\la}\lp c^I D_\om b_I + \half e^I(b\times b)_I 
+  \dfrac{\s\LA}{2} e^I(c\times c)_I \rp\rp
\]
where\footnote{We use the notation 
$(X\times Y)^I \equiv \e_{JK}{}^I X^J Y^J$.} 
\[
F^I(\om)= d\om^I + \half(\om\times\om)^I\,,\quad
D_\om b^I = db^I + (\om\times b)^I\,.
\]
The action $S_2$ describes a pair of 1-form "topological matter" fields 
$b^I,\,c^I$ coupled to a first order gravitation theory described 
by the spin connection $\om^I$ and the dreibein $e^I$. $\LA$ is the 
cosmological constant and $\la$ a coupling constant. With the 
redefinitions $b'=\sqrt{|{\la}|}b$
and $c'=\sqrt{|{\la}|}c$ and taking the limit $\LA={\la}=0$ one 
recovers the ``BCEA'' action 
of~\cite{carlip-gegenberg,carlip-gegenberg-mann,freidel-mann-popescu} 
as a special case. 
However, and as it has already been noted by these authors in their 
particular case, 
the general case considered in the present paper may lead to 
equivalent  interpretations where the roles of 
$e^I$, $b^I$ and $c^I$ as dreibein and matter are permuted. 
These alternatives are
 related to the various possible choices for the signs of the parameters 
$\LA$ and ${\la}$. One sees 
 from the discussion made in the Appendix, and especially looking at the  
Table \ref{tableA1}, that this also corresponds to permutations of the 
roles of the parameters $\LA$ and $\la$ as cosmological and coupling constants.

Now, since non-vanishing $\LA$ and $\la$ imply the existence of four non-degenerate invariant quadratic forms, one has to consider the general action \equ{gen-action-JPQR}. However, as it stands, this action would lead to a rather complicated and non-practical  formulation. Substantial simplification arises if one uses the factorization property explained in the Subsection \ref{factorization} of the Appendix. 

We concentrate from now on to the case of Lorentzian signature $\s=-1$ and positive parameters $\LA$ and $\la$:
\eq 
\mbox{signs } (\s,\LA,\la) = (-,+,+)\,,
\eqn{sads-param}
corresponding to the first line in the Tables \ref{tableA1}, \ref{tableA2} and \ref{tableA3} of the Appendix.
The cases corresponding to the   first,  second and third lines of the tables are equivalent. 
We don't treat the fourth line's case, where   the factorization is 
not of the form of  \ads$_+$ $\oplus$ \ads$_-$\footnote{  In this case the factorization is so(2,2) $\oplus$ so(2,2), see table \ref{tableA3}. The maximal compact sub-algebra is the Abelian
 u(1)$^{\oplus4}$,
see Table \ref{tableA2}.}, neither the Riemannian ones ($\s$=1).
  Thus, in our case, the algebra \sads factorizes  in two de Sitter sub-algebras  \ds$_\pm$ as shown in 
\equ{A-factorization}. 
Expanding the connection \equ{sads-connection} in the factorized basis 
\equ{generators-ds-pm}, we obtain
\eq\ba{l}
\AA = \AA_+ + \AA_-\,,\quad
\AA_\pm = \om^I_\pm J_{I\pm} + e^I_\pm P_{I\pm}
\equiv \dsum{A=1}{6}T_A \AA_\pm^A\,,
\ea\eqn{conn-factor-basis}
with
\eq 
\om^I_\pm = \om^I \mp \sqrt{\LA \la} c^I\,,\quad
e^I_\pm = \sqrt{\LA} e^I \mp \sqrt{\la} b^I\,.
\eqn{coord-change1}
The action \equ{gen-action-JPQR} is now the sum of two de Sitter Chern-Simons actions
\eq 
S = S_+ + S_- = \ka_+\lp S^{(1)}_+ - \frac{1}{\g_+} S^{(2)}_+\rp
+ \ka_-\lp S^{(1)}_- - \frac{1}{\g_-} S^{(2)}_- \rp\,,
\eqn{factorized-action}
where $\ka_\pm$ and $\g_\pm$ are non-zero finite real 
parameters\footnote{$\g_+$ and $\g_-$ are two analogs of the 
Barbero-Immirzi parameter~\cite{barbero-immirzi} $\g$ in dimension (1+3) loop quantum gravity, which share with it the property of not appearing in the classical field equations. See also~\cite{bonzom-livine} in the context of the dimension (1+2) de Sitter theory.}, and
\eq
S^{(n)}_\pm = -\dint_\MM k^{(n)}_{AB}
\lp \AA_\pm^A\lp d\AA^B_\pm + \frac13(\AA_\pm\times\AA_\pm)^B\rp \rp\,,
\quad n=1,2\,,
\eqn{actions-1-2}
are the actions calculated using the two independent invariant quadratic  
forms~\cite{Witten:1988hc,bonzom-livine, BCOP} $k^{(n)}$ $(n=1,2)$ belonging 
to each of the algebras \ds$_\pm$, as shown in Eqs.  
\equ{2-quad-forms} of the Appendix:
\eq\ba{l}
k^{(1)}_{J_\pm^I,J_\pm^J} = \eta_{IJ}\,,\quad 
k^{(1)}_{P_\pm^I,P_\pm^J} = -\eta_{IJ}\,,\esp
k^{(2)}_{J_\pm^I,P_\pm^J} = \eta_{IJ}\,.
\ea\eqn{2-quad-forms-pm}
(We only write the non-vanishing elements). 

  Each individual action $S_i$ in \equ{gen-action-JPQR} would lead to the same field equations, and therefore the total action $S$ leads to equations independent of the parameters 
  $c_i$ -- in \equ{gen-action-JPQR} --
  or $\ka_\pm,\,\g_\pm$ -- in \equ{factorized-action}. These equations read simply, in the factorized formulation,
 \eq
\FF_\pm = 0\,,\quad\mbox{with}\quad \FF_\pm = d\AA_\pm+\AA_\pm\AA_\pm\,.
\eqn{field-eq-fact} 
 
 With the signs of its parameters given in \equ{sads-param}, the gauge algebra \sads\ 
possesses a compact subalgebra so(4), as seen in Subsection 
\ref{max-compact=subalg} of the Appendix. Its basis generators 
are listed in the first line of Table \ref{tableA1}. 
With the factorization \equ{A-factorization}, 
so(4) correspondingly factorizes as
\eq 
{\rm so(4)} = {\rm so(3)}_+ \oplus {\rm so(3)}_-\,,\quad
\mbox{with\ } {\rm so(3)}_\pm \subset {\rm \ds}_\pm\,.
\eqn{factorization-so(4)}
A convenient new basis of the de Sitter sub-algebras \ds$_\pm$ is given by 
the generators $L_\pm^i$ and $K_\pm^i$ (i=1,2,3), with the $L_\pm^i$'s forming a
basis of the sub-algebra so(3)$_\pm$ of \ds$_\pm$ = so(3,1)$_\pm$, and the
"boosts" 
$K_\pm^i$'s generating the non-compact part of \ds$_\pm$. This new basis
satisfies the commutation rules\footnote{Indices $i,\,j,\,\cdots$ are 
raised and lowered by the Euclidean metric $\d_{ij}$.
It will be convenient to adopt a vector-like notation,
%\eg $A = (A^i,\,i=1,2,3)$,  
$A^{i}B_{i}= A\cdot B$, 
$\e_{ijk}A^j B^k=(A\times B)_i$, etc.}
\[
[L^{i}_\pm,L^{j}_\pm]=\varepsilon^{ij}{}_{k}L^{k}_\pm,\quad
[L^{i}_\pm,K^{j}_\pm]=\varepsilon^{ij}{}_{k}K^{k}_\pm,\quad
[K^{i}_\pm,K^{j}_\pm]=-\varepsilon^{ij}{}_{k}L^{k}_\pm\,, 
\]
and is defined as%\footnote{}
\eq
L_\pm = (P^{2}_\pm/\sqrt{\Lambda},-P^{1}_\pm/\sqrt{\Lambda},J^{0}_\pm)\,, \quad
K_\pm = (J^{2}_\pm,-J^{1}_\pm,-P^{0}_\pm/\sqrt{\Lambda})\,,
\eqn{generators-L-K}
the \ds$_\pm$ generators $J^{I}_\pm$ and $P^{I}_\pm$ being expressed in
terms of the original generators $J^I$, $P^I$, $Q^I$ and $R^I$ by Eq.
\equ{generators-ds-pm}.
The expansion of the \sads\ connection \equ{sads-connection}
in the basis $L_\pm,\,K_\pm)$
reads
\[
\AA = \AA_+ + \AA_-\,,\quad
\AA_\pm = A_\pm \cdot L_\pm + B_\pm \cdot K_\pm\,,
\]
with
\[
A_\pm = (\sqrt{\Lambda}e_\pm^{2},-\sqrt{\Lambda}e_\pm^{1},-\omega_\pm^{0})\,,\quad
B_\pm = (\omega_\pm^{2},-\omega_\pm^{1},\sqrt{\Lambda}e_\pm^{0})\,,
\]
the components $e_\pm^I$ and $\om_\pm^I$ being given in \equ{coord-change1}.

We can now write the action $S_\pm$ in terms of the new field components
as\footnote{Boldface letters
represent space objects, e.g. $\bfA=A_a dx^a$ $=$ $(A^i_a dx^a,\,i=1,2,3)$  , etc.}
\begin{equation}\ba{l}
S_\pm=-\dfrac{\kappa_\pm}{2}\!\!\int_\mathbb{R}{\rm d}t\LP\int_\S
\left(\dot{\bfA}_\pm\cdot(\bfB_\pm-\frac{1}{\gamma_\pm}\bfA_\pm)
+\dot{\bfB}_\pm\cdot(\bfA_\pm
+\frac{1}{\gamma_\pm}\bfB_\pm)\right)\esp
\phantom{S_\pm=-\frac{\kappa_\pm}{2}\!\!\int_\mathbb{R}{\rm d}t\LP}
-\mathcal{G}_\pm(A_{t\pm}) - \mathcal{G}_{0\pm}(B_{t\pm})\RP,
\label{CS:SI-SII}\ea\end{equation}
where
\begin{subequations}
\begin{align}
\mathcal{G}_\pm(A_{t_\pm}) 
= \kappa_\pm\int_\S A^i_{t\pm}(\bfx)\GG^i_{\pm}(\bfx)
& =\kappa_\pm\int_\S A_{t\pm}\cdot[{\bfD}\bfB_\pm
-\frac{1}{\gamma_\pm}(\bfF_{\bfA_\pm}-\frac{1}{2}\bfB_\pm\times\bfB_\pm)]\,,
\label{GG}\\
\mathcal{G}_{0\pm}(B_{t\pm}) 
= \kappa_\pm\int_\S B^i_{t\pm}(\bfx)\GG^i_{0\pm}(\bfx)
& =\kappa_\pm\int_\S B_{t\pm}\cdot[\bfF_{\bfA_\pm}
-\frac{1}{2}\bfB_\pm\times\bfB_\pm+\frac{1}{\gamma_\pm}\bfD\bfB_\pm],
\label{GG_0}\end{align}
 \end{subequations} 
with $\bfF_{\bfA_\pm}=\mathbf{d}\bfA_\pm+\frac{1}{2}\bfA_\pm \times\bfA_\pm$ and
$\bfD\bfB_\pm=\mathbf{d}\bfB_\pm+\bfA_\pm\times\bfB_\pm$.

 One first note that the conjugate momenta 
of $A^i_{t\pm}$ and $B^i_{t\pm}$ are primary constraints, in
Dirac's terminology~\cite{dirac}, whereas $\GG_\pm(A_{t\pm})$ and $\GG_{0\pm}(B_{t\pm})$
are the secondary constraints ensuring the stability of the former primary constraints. 
$A^i_{t\pm}$ and $B^i_{t\pm}$ then play the role of
Lagrange multipliers. There are still two primary constraints involving the conjugate
momenta of the fields $A^i_{a\pm}$ and $B^i_{a\pm}$. They turn out to be of
second class, whose solution according to the Dirac-Bergmann
algorithm~\cite{dirac} gives rise to the Dirac-Poisson brackets
\begin{align}
\{A_{a\pm}^{i}(\bfx),A_{b\pm}^{j}(\bfx')\} & =
\frac{1}{\kappa_\pm}\e_{ab}\delta^{ij}\frac{\gamma_\pm}{1+\gamma_\pm^{2}}
\delta^{2}(\bfx-\bfx'),\nonumber \\
\{B_{a\pm}^{i}(\bfx),A_{b\pm}^{j}(\bfx')\} & =
-\frac{1}{\kappa_\pm}\e_{ab}\delta^{ij}\frac{\gamma\pm^{2}}{1+\gamma_\pm^{2}}
\delta^{2}(\bfx-\bfx')\label{Sym.2},\\
\{B_{a\pm}^{i}(\bfx),B_{b\pm}^{j}(\bfx')\} & =
-\frac{1}{\kappa}\e_{ab}\delta^{ij}\frac{\gamma_\pm}{1+\gamma_\pm^{2}}
\delta^{2}(\bfx-\bfx').\nonumber 
\end{align}
The resulting Hamiltonian turns out to be fully constrained, as it is
expected in such a general covariant theory:
\begin{equation}
H= H_+ + H_-\,,\quad
H_\pm = \mathcal{G}_\pm (A_{t\pm})+\mathcal{G}_{0\pm}(B_{t\pm}),
\label{Hamil}\end{equation}
with the constraints $\GG_\pm$ and $\GG_{0\pm}$ as given by \equ{GG}, \equ{GG_0}.
These constraints are first class, obey the Dirac-Poisson bracket
algebra (we only write the non-zero brackets)
\begin{align}
\{\mathcal{G}_\pm(\varepsilon),\mathcal{G}_\pm(\varepsilon')\} & 
=\mathcal{G}_\pm(\varepsilon\times\varepsilon')\,,\nonumber\\
\{\mathcal{G}_{0\pm}(\varepsilon),\mathcal{G}_\pm(\varepsilon')\} & 
=\mathcal{G}_{0\pm}(\varepsilon\times\varepsilon')\,,\\
\{\mathcal{G}_{0\pm}(\varepsilon),\mathcal{G}_{0\pm}(\varepsilon')\} & 
=\sigma\mathcal{G}_\pm(\varepsilon\times\varepsilon')\nonumber\,, 
\end{align}
and also generate the gauge transformations
under which the theory is invariant:
\begin{equation}
\begin{aligned}
\{\mathcal{G}_\pm(\varepsilon),\bfA_\pm\} & =\bfD_\pm\varepsilon, & 
\{\mathcal{G}_\pm(\varepsilon),\bfB_\pm\} & = \bfB_\pm \times\varepsilon;\\
\{\mathcal{G}_{0\pm}(\varepsilon'),\bfA_\pm\} & =-\bfB_\pm\times\varepsilon' & 
\{\mathcal{G}_{0\pm}(\varepsilon'),\bfB_\pm\} & = \bfD_\pm\varepsilon'
\quad\quad\quad (\bfD_\pm=\bfd+\bfA_\pm\times).
\end{aligned}
\label{Gau:Tr:BF}
\end{equation}
Invariance under these gauge transformations ensures
diffeomorphism invariance, up to field equations. Indeed, infinitesimal
diffeomorphisms, given by the Lie derivative, can be written as 
infinitesimal gauge transformations with parameters 
$(\varepsilon,\varepsilon')=(\imath_{\xi}\bfA_\pm,\imath_{\xi}\bfB_\pm)$, 
up to field equations:
\begin{equation}
\begin{array}{cc}
\pounds_{\xi}\bfA_\m & = \bfD_\pm\e - B_\pm\times \e' 
+\ \mbox{field equations,}\\
\pounds_{\xi}\bfB_\pm & =\bfD_\pm\e' + \bfB_\pm\times\e 
+\ \mbox{field equations,}
\end{array}
\label{difeo}\end{equation}
with $\pounds=\mathrm{d}\imath_{\xi}+\imath_{\xi}\mathrm{d}$ the
Lie derivative. 

%%%%%%%%%%%%%%%%%%%%%%%%%%%%%%%%%%%%%%%%%%%%%%%%%%%%%%%%%%%%%%%%
\section{Partial gauge fixing: the axial gauge}\label{calibre-axial}

In order to be left with a compact gauge group, we partially fix the
gauge, fixing the "boost" gauge degrees of freedom, which correspond to
the gauge transformations generated by the generators $K^i_\pm$ defined
in \equ{generators-L-K}. This   is  done imposing new constraints
$B^i_{y\pm}\approx0$,
implemented by the addition of the terms 
\[
\dint_\S d^2x\lp\mu_{i+}(\bfx) B^i_{y+}(\bfx)  +
\mu_{i-}(\bfx) B^i_{y-}(\bfx)\rp
\]
to the Hamiltonian \equ{Hamil}, with $\mu_{i\pm}$ as Lagrange multiplyer fields.
The 6 gauge fixing constraints together with the 6 constraints
$\GG^i_{0\pm}(\bfx)$ defined in \equ{GG_0} are second class, hence
become strong equalities through Dirac's redefinition of the brackets. 
After insertion of the gauge fixing constraints, the $\GG_0$
constraints read
\[
\partial_{x}A^i_{y\pm}-D_{y\pm}\lp A^i_{x\pm}+\frac{1}{\gamma_\pm}B^i_{x}\rp =0 \,,
\]
and can be solved  for $B_{x\pm}^i$ as
functionals of $A^i_{x\pm}$ and $A^i_{y\pm}$. The number of independent dynamical
fields is now reduced to 12, which can be conveniently chosen as
\[
\mathcal{A}^i_{x\pm} = A^i_{x\pm}-\gamma_\pm B^i_{x\pm} ,\quad
\mathcal{A}^i_{y\pm} = A^i_{y\pm} \,,
\]
obeying the Dirac-Poisson algebra
\eq
\{\mathcal{A}^i_{x\pm}(\bfx),\,\mathcal{A}^j_{y\pm}(\bfx')\}_{\rm D} =
\frac{\gamma_\pm}{\kappa_\pm}\d^{ij}\d^2(\bfx-\bfx') \,,
\eqn{DB-cal-A}
where $\{\ \,,\ \}_{\rm D}$ denotes the Dirac bracket.
In these variables, the Hamiltonian reads
\eq
H = H_+ + H_- \,,\quad H_\pm = -\dfrac{\kappa_\pm}{\g_\pm}\GG_\pm(A_{t\pm}) \,,
\eqn{CS-hamiltonian}
with the first class constraint  $\GG_\pm$ given by
\eq
\GG_\pm(\eta) = \dint_\S d^2x
\lp \eta_{i+}(\bfx)\FF_+^i(\bfx) + \eta_{i-}(\bfx)\FF_-^i(\bfx) \rp \approx0 \,,
\eqn{curv-constr}
or equivalently by the curvature constraints
\eq
\FF_\pm^i(\bfx) \equiv 
\partial_{x}\mathcal{A}_{y}^{i}-\partial_{y}\mathcal{A}_{x}^{i}
+{\e}^i{}_{jk}\mathcal{A}_{x}^{j}\mathcal{A}_{y}^{k}
\approx 0\,.
\eqn{curv-constr'}
The basic Dirac-Poisson brackets \equ{DB-cal-A}, together with the expressions 
\equ{CS-hamiltonian}, \equ{curv-constr} for the Hamiltonian 
show that the theory is reduced to a
Chern-Simons theory for the two so(3) connections $\bfAA_\pm$, which indeed 
transform as
\[
\{\GG_\pm(\eta_\pm),\,\bfAA^i_{a\pm}\}_{\rm D} = 
\pa_a\eta_\pm^i + \e^i{}_{jk}\bfAA^j_{a\pm} \eta_\pm^k ,
\]
under the gauge transformations induced by the constraints.

We can summarize the result saying that we have a Chern-Simons theory for the 
so(4) connection\footnote{$\a,\,\b = 1, \cdots,6$ are so(4) = so(3)$_+$ $\oplus$ so(3)$_-$
indices, whereas i,\,j =1,2,3 are so(3)$_\pm$ ones.}
\[
\bfAA \equiv \dsum{\a=1}{6} \bfAA^\a \TT_\a = 
\bfAA_++\bfAA_-\,, \quad \bfAA_\pm = \bfAA^i_a T_{i\pm} dx^a\,,
\]
with the constraints \equ{curv-constr'}, which may be written as
\[
\FF \equiv \FF_+ + \FF_- \approx0\,.
\]
The   basis $(\tau_\a,\,\a=1,\cdots,6)$ = 
$( \, T_{i+} ,\,T_{i-},\,i=1,2,3)$  for the algebra so(4) obeys the commutation relations 
\[
[T_{i\pm},T_{j\pm}] = \e_{ij}{}^k T_{k\pm}\,,\quad
[T_{i+},T_{j-}] = 0\,.
\]
For further use, we normalize the Killing forms 
of so(4) and so(3)$_\pm$, denoted by the symbol Tr, as
\[
{\rm Tr} (\XX\YY) \equiv \sum_{\a=1}^6\XX^\a \YY^\a\,,\ \XX,\YY\in {\rm so}(4)\,;\quad
{\rm Tr} (XY) \equiv \sum_{i=1}^3 X^i Y^i\,,\ X,Y\in {\rm so}(3)_\pm\,.
\]

%%%%%%%%%%%%%%%%%%%%%%%%%%%%%%%%%%%%%%%%%%%%%%%%%%%%%%%%%%%%
\section{ Quantization}\label{quantization}

We apply to the present model the quantization procedure followed 
in~\cite{Constantinidis-Luchini-Piguet,BCOP}. We have first
to choose the gauge group since we will have to go from the Lie
algebra level to the group level. 
Since the residual gauge invariance left after the partial gauge 
fixing made in the preceding Section is so(4), a convenient 
choice\footnote{In D=4 LQG, where the choice for the gauge group is SU(2), and not SO(3), which allows the coupling with fermions. The motivation for our present choice of Spin(4), and not SO(4), is similar, although its physical necessity is not as strong.} 
is the
universal covering of SO(4), namely Spin(4) = SU(2)$\times$SU(2).

The dynamical field
variables $\bfAA^i_{a\pm}$, components of the so(4) connection $\bfAA$
defined after the partial gauge fixing, 
are taken now as operators obeying the
commutation rules (we display only the non-vanishing commutators)
\begin{eqnarray}
[\hat{\mathcal{A}}_{x\pm}^{i}(\bfx),\hat{\mathcal{A}}_{y\pm}^{j}(\bfx')]
=\frac{i\gamma_\pm}{\kappa_\pm}\,\delta^{ij}\delta^{2}(\bfx-\bfx'),
\end{eqnarray}
where  $i,j=1,2,3$ are  the so(3) indices. The task is to find a
representation of this algebra in some kinematical Hilbert space, and 
then to apply the constraints. We shall therefore consider a space of
wave functionals $\Psi[\bfAA_x]$ =  $\Psi[\bfAA_{x+},\bfAA_{x-}]$ 
where the conjugate variables 
$\bfAA_{y\pm}$ act as functional derivatives:
\[
\bfAA^i_{y\pm}(\bfx)\Psi[\bfAA_x] = \dfrac{\g_\pm}{i\kappa_\pm}
\dfud{}{\bfAA^i_{x\pm}(\bfx)}\Psi[\bfAA_x] \,.
\]
The quantum version of the curvature constraints \equ{curv-constr'} read
\begin{eqnarray}
\left(i\left(\partial_{x}\frac{\delta}{\delta \mathcal{A}_{x}^{i\pm}}
+f^{i}{}_{jk}\mathcal{A}_{x\pm}^{j}\frac{\delta}{\delta \mathcal{A}_{x\pm}^{k}}\right)
+\dfrac{\kappa_\pm}{\g_\pm}\partial_{y}\mathcal{A}_{x\pm}^{i}\right)
\Psi[\bfAA_x]
=0\,,
\label{gauss_c}\end{eqnarray}
and a particular solution is given by~\cite{quantiz-of-CS} 
\eq
\Psi_0[\bfAA_x]=\exp(2\pi i\alpha_{0+}) \exp(2\pi i\alpha_{0-})\,,
\eqn{sol-Psi_0}
with 
\eq\ba{l}
\alpha_{0\pm}=\frac{\kappa_\pm}{6\pi\g_\pm}
\dint_{\tilde{\Sigma}}d^{3}x\,\epsilon^{\mu\nu\rho}{\rm Tr}
(h_\pm^{-1}\partial_{\mu}h_\pm\, h_\pm^{-1}\partial_{\nu}h_\pm\, 
h_\pm^{-1}\partial_{\rho}h_\pm) \esp
\phantom{\alpha_{0\pm}=}
-\frac{\kappa_\pm}{2\pi\g_\pm}
\dint_{ \Sigma=\partial {\tilde  \Sigma} }d^{2}x\,
{\rm Tr}(\mathcal{A}_{x\pm}h_\pm^{-1}\partial_{y}h_\pm)\,,
\ea\eqn{alpha_0}
where $h_\pm(\bfx)$ is an element of the gauge group SU(2)$_\pm$,
defined  as a functional of $\mathcal{A}_{x\pm}$ by 
\eq
\mathcal{A}_{x\pm}=h_\pm^{-1}\partial_{x}h_\pm \,,
\eqn{def-h}
and where $\tilde\S$ is a 3-manifold having the space sheet $\S$ as its
 border. 
The first term in \equ{alpha_0} is the Wess-Zumino-Witten action. 
 The group being 
non-abelian and compact, the integral over $\tilde\S$ is defined 
up to the addition of a constant 
$24\pi^2\,\n_\pm$, with $\n_\pm\in\mathbb{Z}$. This requires
that each ratio $\kappa_\pm/\g_\pm$ must be quantized~\cite{witten-WZW}:  
\eq
\dfrac{\kappa_\pm}{\g_\pm}=\dfrac{\nu_\pm}{4\pi}\,.
\eqn{quant-cond}
The general solution of the constraints then can be written as
\eq
\Psi[\bfAA_x]=\Psi_{0}[\bfAA_x]\psi'[\bfAA_x],
\eqn{Psi-Psi0}
where the reduced wave functional $\psi'[\bfAA_x]$  satisfies
\eq
\left[i\left(\partial_{x}\frac{\delta}{\delta \mathcal{A}_{x\pm}^{i}}
+f^{i}\!_{jk}\mathcal{A}_{x\pm}^{j}
\frac{\delta}{\delta \mathcal{A}_{x\pm}^{k}}\right)\right]
\psi'[\bfAA_x]=0\,.
\eqn{curv-constraints''}
The latter equations mean that $\psi'$ is invariant under the  
infinitesimal ``$x$-gauge  transformations''
\eq
\delta \mathcal{A}_{x\pm}^{i}=D_{x\pm}\epsilon_\pm^{i} \,.
\eqn{x-gauge-transf}
Following the general lines of loop quantization~\cite{general-ref}, 
we introduce holonomies of the so(4) connection component $\bfAA_x$
 as configuration space variables, the reduced wave functionals $\psi{\rm ^{inv}}$
being then functions of them. 
As in~\cite{Constantinidis-Luchini-Piguet,BCOP} we take as the space
sheet $\S$ a space having the topology of a cylinder, for which we
choose coordinates $x,\,y$ with $0\leq x\leq 2\pi$ and $-\infty<y<+\infty$.
The holonomies are thus defined along oriented paths $c(y)$ at constant $y$: 
\[
U(y)=\mathcal{P}\exp{\int_{c(y)}\mathcal{A}_{x}dx} = U_+(y)U_-(y)\,,\quad
U_\pm(y)=\mathcal{P}\exp{\int_{c(y)}\mathcal{A}_{x\pm}^{i}(x,y) T_{i\pm}dx}\,,
\]
where $\PP$ means path ordering.
Anticipating the
requirement of the wave functionals 
having to satisfy the constraints \equ{curv-constraints''},
which is
equivalent to require the invariance under the $x$-gauge
transformations \equ{x-gauge-transf}, we shall restrict ourselves to cycles, i.e.,
to paths $c(y)$ which
are closed. If the cycle $c(y)$ begins and ends at the point $(x,y)$, the holonomy transforms as
\eq
U(y)\mapsto g^{-1}(x,y)U(y)g(x,y)\,,\quad
U_\pm(y)\mapsto g_\pm^{-1}(x,y)U_\pm(y)g_\pm(x,y)\,,
\eqn{holotransf}
where $g=g_+g_-\in$ Spin(4) and $g_\pm\in$ SU(2)$_\pm$
We now define the vector space Cyl as the set of ``cylindrical'' wave functionals, 
defined as arbitrary finite linear combinations of wave functionals of the form
\[\ba{l}
\Psi_{\Gamma,f}[\bfAA_x] = 
\Psi_0[\bfAA_x]\psi'_{\Gamma,f}[\bfAA_x]\,,\quad
\mbox{with}\quad \psi'_{\Gamma,f}[\bfAA_x]
= f(U(y_{1}),...,U(y_{K}))\,,
\ea\]
for arbitrary $K$ and arbitrary ``graphs'' $\GA$ defined as finite 
sets of $K$ cycles $c(y_k)$ (see Figure). 

%%%%%%%%%%%%%%%%%%%%%%%%%%%%%%%%%%%%%%%%%%%%%%%%%%%%%
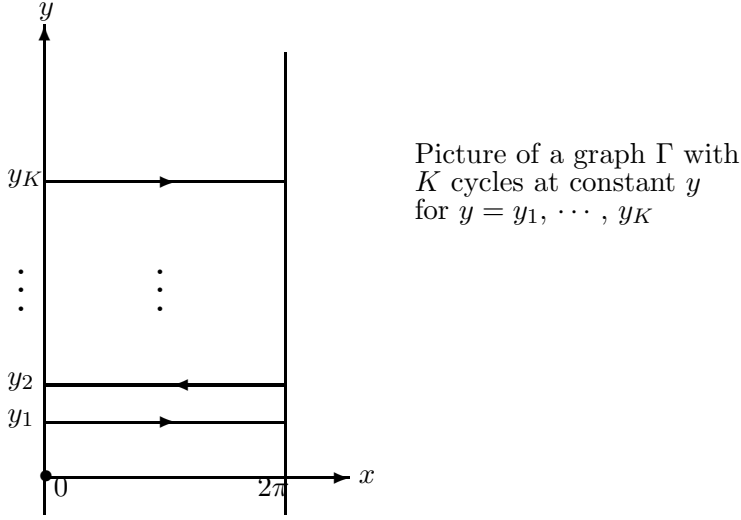
\begin{picture}(250,270)(0,0)
\setlength{\unitlength}{.7pt}
%\setlength{\unitlength}{1.0pt}
%\qbezier(100,0)(150,70)(250,75)
\thicklines
%\put(140,100){\circle*{2}}
\put(200,40){\vector(1,0){15}}
\put(50,270){\vector(0,1){15}}
\put(50,40){\line(1,0){150}} %x-axis
\put(50,20){\line(0,1){250}} %y-axis
\put(47,37){$\bullet$}
\put(55,30){$0$} \put(165,30){$2\pi$} \put(220,37){$x$}
\put(47,290){$y$}
\put(180,20){\line(0,1){250}} % // y-axis
% Cycles:
\put(30,70){$y_1$}
\put(50,70){\line(1,0){130}} \put(120,70){\vector(1,0){1}}
%%%%%%%%%%%%%%%%%%%%%%%%
\put(30,90){$y_2$}
\put(50,90){\line(1,0){130}} \put(120,90){\vector(-1,0){1}}
%%%%%%%%%%%%%%%%%%%%%%%%
 \put(35,130){{\bf .}} \put(35,140){{\bf .}} \put(35,150){{\bf .}}
 \put(110,130){{\bf .}} \put(110,140){{\bf .}} \put(110,150){{\bf .}}
%%%%%%%%%%%%%%%%%%%%%%%%
\put(30,200){$y_K$}
\put(50,200){\line(1,0){130}} \put(120,200){\vector(1,0){1}}
%%%%%%%%%%%%%%%%%%%%%%%%
\put(250,210){Picture of a graph $\GA$ with} 
\put(250,195){$K$ cycles at constant $y$}
\put(250,180){for $y=y_1,\,\cdots,\,y_K$}
\label{Cilindro}
\end{picture}
%%%%%%%%%%%%%%%%%%%%%%%%
%\vspace{-140pt}

Since the cylindrical functionals are functions Spin(4)$\,\otimes\cdots\otimes\,$Spin(4)
$\to$ $\complex$, a scalar product can be defined using the Spin(4) invariant Haar
measure\footnote{We use Dirac's notation, with $\vev{\bfAA_x|\GA,f}$ =
$\Psi_{\GA,f}[\bfAA_x]$}: 
\[
\vev{\GA,f|\GA',f'} = \dint \lp\prod_{k=1}^{\tilde K} d\m(g_k)\rp
\lp f(g_1,\cdots,g_K\rp^* f'(g'_1,\cdots,g'_{K'})\,,
\]
with $d\m(g\!=\!g_+g_-)$ = $d\m_+(g_+)d\m_-(g_-)$ the normalized Haar measure 
of Spin(4) and $\tilde\GA$ the union 
of the graphs $\GA$ and $\GA'$.
This internal product allows us to define the non-separable Hilbert space $\overline{\rm Cyl}$ as
the Cauchy completion of Cyl.

An orthonormal basis of $\overline{\rm Cyl}$, the spin network basis, is provided, 
thanks to
Peter-Weyl's theorem, by the wave functionals
\eq
\Psi_{\Gamma,\vec j,\,\vec\a,\,\vec\b}[\bfAA_x] = 
\Psi_0[\bfAA_x] \prod_{k=1}^{K}\sqrt{2j_k^++1}\, R^{j_k^+}_{\a_k^+,\,\b_k^+}\, \sqrt{2j_k^-+1}\,R^{j_k^-}_{\a_k^-,\,\b_k^-}\,,
\eqn{Cyl-basis}
where we have associated to each cycle $c(y_k)$ the matrix elements of a unitary irreducible representation of Spin(4), labelled by the half-integer spin pairs $(j_k^+,\,j_k^-)$,
of the corresponding holonomies: 
\eq
R^{j_k^\pm}_{\a_k^\pm,\,\b_k^\pm} 
= \lp R^{j_k^\pm}\lp U_\pm(y_k)\rp \rp_{\a_k^\pm,\,\b_k^\pm}\,.
\eqn{repr-matrix}
The representation $(0,\,0)$, if present for some cycle $c(y_k)$, would yield a vector already present in the set of basis vectors corresponding to the graph obtained from $\GA$ by deleting this cycle.
In order to avoid redundancy, we therefore exclude such a representation.
The orthonormality conditions read
\[
\vev{\Gamma,\vec j,\,\vec\a,\,\vec\b | \Gamma',{\vec j}',\,{\vec\a}',\,{\vec\b}'} = \d_{\GA\GA'}  \d_{{\vec j}{\vec j}'} \d_{{\vec\a}{\vec\a}'} \d_{{\vec\b}{\vec\b}'}\,.
\]
The curvature constraint in the form \equ{curv-constraints''} is readily implemented by taking the traces of the representation matrices 
\equ{repr-matrix}, the characters $\chi^{j_k^+,\,j_k^-}$
= $\trace\lp R^{j_k^+}\rp\trace\lp R^{j_k^-}\rp$. We define in this way the Hilbert subspace $\HH_{\rm kin}$ of $\overline{\rm Cyl}$, of orthonormal basis
\[
|\GA,\vec j\rangle\,,\quad \vev{\GA,\vec j|\GA',{\vec j}'}=\d_{\GA\GA'}  \d_{{\vec j}{\vec j}'}\,,
\]
with
\[
\vev{\bfAA_x|\GA,\vec j} = \prod_{k=1}^{K}
\sqrt{2j_k^++1}\, \chi^{j_k^+}\, \sqrt{2j_k^-+1}\,\chi^{j_k^-}\,.
\]
$\HH_{\rm kin}$ is still non-separable since each vector of its orthonormal basis depends on a set of real numbers $y_k$ characterizing each graph $\GA$.
This defect is due to our particular choice for the class of coordinates 
$x,\,y$ adapted to the cylinder's topology. Invariance under general transformations of the $y$-coordinate -- the ``$y$-diffeomorphisms'' 
  in the point of view of active transformations --  
is not yet fulfilled. In order to implement it, a group averaging over the group of $y$-diffeomorphisms has to be performed, with the result that two basis vectors $|\GA,\vec j\rangle$ and 
$|\GA',\vec j\rangle$
 corresponding to two graphs which are related to each other by a $y$-diffeomorphism but sharing the same spin labels, represent the same physical vector 
\eq 
|\vec j\rangle = |j_1^+,\cdots,j_K^+,\,j_1^-,\cdots,j_K^-\rangle\,, 
\eqn{H-phys-basis}
element of the orthonormal basis of the physical Hilbert space  
 $\HH_{\rm phys}$. Obviously, the latter Hilbert space is 
 separable\footnote{See \cite{Constantinidis-Luchini-Piguet,BCOP} for more details.}.

%%%%%%%%%%%%%%%%% wess zumino%%%%%%%%%%%%%%%%%%%%%%%%%%%%%%%%%%%%%%%%%%%
\section{Observables}\label{observavel}
A pair of observables $L\pm$ which are diagonal in the spin basis 
\equ{H-phys-basis} of
$\HH_{\rm phys}$ can be constructed following the lines of~\cite{BCOP}.

At the classical level, they are given by the expressions
\[
L_\pm( b) = \dint_{\!\!\!\!b} dy \sqrt{\sum_{i=1}^{3} W^i_{y\pm}W^i_{y\pm}}
= \dint_{\!\!\!\!b} dy\sqrt{\trace W^2_{y\pm}}\,,
\]
where $ b$ is an infinite curve $\{-\infty<y<\infty\}$ at constant $x$, and
\[
W_{y\pm} = \bfAA_y - h_\pm^{-1}\pa_y h_\pm\,,
\]
with $h\pm$ given as a non-local functional of $\bfAA_\pm$ as a solution of \equ{def-h}. As $h_\pm$ transforms as $h'_\pm$ = $h_\pm g_\pm$ under a 
SU(2)$_\pm$ gauge transformation $g_\pm$, 
  the expression  $h_\pm^{-1}\pa_y h_\pm$ transforms as a
 connection, hence $\hat W_{y\pm}$ is in the adjoint representation and $L_\pm( b)$ is gauge  invariant: the latter are candidates for observables. It turns out that, as shown in~\cite{BCOP}, the classical $W_{y\pm}$, hence $L_\pm( b)$, are vanishing. However their quantum counterparts are not,   as we show now. 

The quantum version of $W_{y\pm}$ reads
\[
\hat W_{y\pm} = \hat\bfAA_y - h_\pm^{-1}\pa_y h_\pm\,.
\]
In order to give a reliable definition of $\hat L_\pm( b)$  as quantum operators in the physical Hilbert space, one first introduces a regularization, analogous to the one used to define the area operator of loop quantum gravity~\cite{general-ref}. One begins the construction in the space $\overline{\rm Cyl}$, then extends it to the kinematical Hilbert space $\HH_{\rm kin}$ and finally to the physical Hilbert space 
$\HH_{\rm phys}$.
The regularization consists first in dividing the integration interval $ b$ in pieces $ b_n$, small enough for each of them to intersect at most one of the cycles of the graph associated to the basis vector 
$|\GA,\,\vec j,\,\vec\a,\,\vec\b\rangle$
of $\overline{\rm Cyl}$ on which $\hat W_{y\pm}$ acts. Second, one defines the operator $\hat L_\pm( b)$  as the sum
\[
\hat L_\pm( b) = \dsum{n}{} \hat L_\pm( b_n)\,,
\]
where $\hat L_\pm( b_n)$ is approximated by 
\[
\hat L_\pm( b_n) = \sqrt{\dsum{i=1}{3} 
\dint_{\!\!\!\!b_n} \hat W^i_{y_\pm} \dint_{\!\!\!\!b_n} \hat W^i_{y_\pm}
}\,.
\]
The result,
\eq
\hat L_\pm(b) |\GA,\,\vec j,\,\vec\a,\,\vec\b\rangle =
\dfrac{\g_\pm}{\ka_\pm} \dsum{k=1}{K} \sqrt{j_k^\pm(j_k^\pm+1)}
|\GA,\,\vec j,\,\vec\a,\,\vec\b\rangle 
\eqn{eigenvalues}
where the summation is performed on all cycles of the graph $\GA$, is independent of further refinements of the partition $b$
=   $\cup_k b_k$.  It is also independent of the location $x$ of the curve $b$. It only depends on the spins associated to each cocycle of the graph $\GA$, independently of its location $y$. This result can therefore be extended to 
$\HH_{\rm kin}$ and then to the physical Hilbert 
space\footnote{See \cite{Constantinidis-Luchini-Piguet,BCOP} for more details.}:
\[
\forall\, |\vec j\rangle \in \HH_{\rm phys}\,,\quad
\hat L_\pm |\vec j\rangle =  \dfrac{4\pi}{\n_\pm}\,
\dsum{k=1}{K} \sqrt{j_k^\pm(j_k^\pm+1)} \,|\vec j\rangle\,,
\] 
where we have used the quantization conditions \equ{quant-cond}, $\n_\pm$ being integers.

%%%%%%%%%%%%%%%%%%%%%%%%%%%%%%%%%%%%%%%%%%%%%%%%%%%%%%%%%%%%%%%
\section{Conclusion}

We have proceeded to the loop quantization of $D=1+2$ gravity with a cosmological constant and a coupling with topological matter fields 
defined via a semi-simple extension of the  de Sitter or anti-De Sitter group. The resulting theory has four free real parameters corresponding to the four different non-degenerate quadratic forms which may be used to construct an action.
But the quantum theory depends only on two independent parameter's ratios, and in fact on 2 integers, $\n_+$ and $\n_-$, due to a topological quantization condition.
An orthonormal spin-network basis has been constructed, the basis vectors being the eigenvectors of two global observables with eigenvalues very similar to those of the area operator in (1+3) - dimensional LQG. 
   These observables are a pure quantum effect, their classical counterparts being vanishing.

%%%%%%%%%%%%%%%%%%%%%%%%%%%%%%%%%%%%%%%%%%%%%%%%%%%%%%%%%%%%%
\appendix
\setcounter{section}{0}

\section*{Appendix}
%%%%%%%%%%%%%%%%%%%%%%%%%%%%%%%%%%%%%%%%%%%%%%%%%%%%%%%%%%%%%%%%%%%
\section{The algebra \sads} 
%%%%%%%%%%%%%%%%%%%%%%%%%%%%%%%%%%%%%%%%%%%%%%%%%%%%%
\subsection{Semi-simple extension of a semi-simple Lie algebra}\label{semi simple ext}
Let $\GG$ be the Lie algebra of a semi-simple Lie group $G$, and 
$\{T_\a, \a=1,\dots,d\}$ a basis of $\GG$, with the commutation relations 
\eq
[T_\a,T_\b] = f_{\a\b}{}^\g T_\g\,.
\eqn{alg-GG}
Let us consider a set of operators $\{S_A, A=1,\dots,D\}$ in a dimension $D$ representation of $\GG$, \ie, transforming as 
\[ 
[S_B,T_\a] = R_{\a B}{}^C S_C
\]
under the action of the basis generators of $\GG$, $R_{\a B}{}^C$ being the elements of the matrix representing $T_\a$.

We define the \textit{semi-simple extension} $S\GG$ of $\GG$ through the representation $R_\a$ the algebra spanned by the operators $\{T_\a,S_A\}$, whereby the commutation relations above are completed by
\[
[S_A,S_B]=C_{AB}{}^\g T_\g\,.
\]
A necessary and sufficient condition for this extension to exist is the fulfilment of the Jacobi identities involving the new structure constants 
%$R_{\a B}{}^C$ and 
$C_{AB}{}^\g$:
\[\ba{l}
%R_{\a C}{}^D R_{\b D}{}^E - R_{\b C}{}^D R_{\a D}{}^E
%- f_{\a\b}{}^\g R_{\g C}{}^E = 0\,,\esp
f_{\a\d}{}^\e C_{BC}{}^\d + R_{\a C}{}^D C_{BD}{}^\e
+ R_{\a B}{}^D C_{D C}{}^\d = 0\,,\esp
C_{AB}{}^\d R_{\d C}{}^E + C_{BC}{}^\d R_{\d A}{}^E 
+ C_{CA}{}^\d R_{\d B}{}^E = 0\,.
\ea\]
The first equation states that $C_{AB}{}^\g$ must an invariant mixed tensor, whereas the second one is a cocycle condition it must fulfils.

A special case is provided by
$S$ = $\{S_A, A=\a=1,\cdots,d\}$ being a vector in the adjoint representation. 
Then $R_{\a B}{}^C=f_{\a\b}{}^\g$ and the cocycle condition is obviously fulfilled by 
$C_{A B}{}^\g= C f_{\a\b}{}^\g$ with $C$ an arbitrary real number. This will be the case of interest in  the present paper, the full commutator  algebra for the basis generators of $S\GG$ being summarized by
\eq 
[T_\a,T_\b] = f_{\a\b}{}^\g T_\g\,,\quad
[S_\a,T_\b] = f_{\a\b}{}^\g T_\g\,,\quad
[S_\a,S_\b] = C f_{\a\b}{}^\g T_\g\,.
\eqn{SGG-algebra}
One notes that, if $C>0$, then the algebra factorizes as 
$S\GG = \GG^+ + \GG^-$, the generators of each factor  $\GG^\pm$  being defined by
\eq 
T^\pm_\a = \half \lp T_\a \pm \frac{1}{\sqrt{C}} S_\a \rp\,,
\eqn{SGG-factorized-alg}
and obeying the same commutation rules as in \equ{alg-GG}.

%%%%%%%%%%%%%%%%%%%%%%%%%%%%%%%%%%%%%%%%%%%
\subsection{Properties of the semi-simple extension of \ads}

The \ads\ algebra being given by the commutation rules written in the first line of \equ{SADS-algebra} for the basis generators $\bsJ,\bsP$, its semi-simple extension \sads\ through the adjoint representation vector $\bsQ,\bsR$ is defined by \equ{SGG-algebra}, the result being given by the full 
system of commutators \equ{SADS-algebra}. The second and third lines of \equ{SADS-algebra} correspond to the second equation of \equ{SGG-algebra}, whereas the last line of \equ{SADS-algebra} represents the cocycle condition given by the third equation of \equ{SGG-algebra}. The closure parameter $C$ is now represented by the parameter ${\la}$ (multiplied by the signature $\s$) appearing in the last line of \equ{SADS-algebra}\footnote{One notes that the \ads\ algebra itself, spanned by the generators $\bsJ,\bsP$, is the semi-simple extension of the Lorentz algebra through the "translation" vector $\bsP$, the closure parameter being the cosmological constant $\LA$.}.

%%%%%%%%%%%%%%%%%%%%%%%%%%%%%%%%%%%%%%%%%%%%%%%%%%%%
\subsubsection{Maximal compact sub-algebras}\label{max-compact=subalg}
For the purpose of the gauge fixing proposed in the main text, we are 
interested in finding the maximal compact sub-algebras of \sads. 
These sub-algebras are spanned by subsets L of the 12 basis 
generators, such that their Killing forms 
  are positive or negative definite. 

Ordering the generators of \sads\ as
\[
\{\TT_\a, \a=1,\dots,12\} = 
\{ J^0,J^1,J^2;P^0,P^1,P^2;Q^0,Q^1,Q^2;R^0,R^1,R^2\}\,,
\]
and writing their commutation relations as
$
[\TT_\a,\TT_\b] = F_{\a\b}{}^\g\, \TT_\g\,,
$
the Killing form $K$ reads
\eq 
K_{\a\b} = -\frac{\s}{2}  F_{\a\g}{}^\d F_{\b\d}{}^\g
= {\rm diag}\,(\s,1,1;\ \LA,\s\LA,\s\LA;\ {\la},\s {\la},\s {\la};\ 
\s\LA {\la}, \LA {\la},\LA {\la})\,.
\eqn{Killing}
  Which are the maximal compact subgroups will depend  on the values of the parameters 
$\s,\LA,{\la}$. For instance, with the signs of $\s,\LA,\la$ being 
$-,+,+ $, the signs of the Killing form eigenvalues are
\[
(-,+,+;\ +,-,-;\ +,-,-;\ -,+,+)\,.
\]
One sees that the 6 negative eigenvalues correspond to the 6 generators
$J^0$, $P^1$, $P^2$; $R^0$, $Q^1$, $Q^2$, which span a compact sub-algebra 
which is easily identified as so(4). Note that the generators 
corresponding to the 6 positive eigenvalues do not span a sub-algebra. 
We conclude that the maximal compact sub-algebra, in this case, is so(4). 
Similar reasoning hold for the other cases. 
The results are displayed in Table~\ref{tableA2}, based on the Killing eigenvalue's signs shown in Table~\ref{tableA1}, for each possible 
choices of signs of $\s,\,\LA,\,\la$.
%%%%%%%%%%%%%%%%%%%%%%%%%%%%%%%%%%%%%%%%%%%%%%%%%%%%%%%%%%%%%%%%%%%%%%%%%%%%
\begin{table}[hbt]
\centering
%\begin{tabular}{|c|c|c|c|c|c|c|c|c|c|c|c|c|}
\begin{tabular}{|c|cccccccccccc|}
\hline
 $\boldsymbol{\mbox{\bf Signs of }\s,\LA,{\la}}$ &$J^0$&$J^1$&$J^2\ $ &$P^0$&$P^1$&$P^2\ $ &$Q^0$&$Q^1$&$Q^2\ $ &$R^0$&$R^1$&$R^2\ $\\ 
\hline%\hline 
$-,\ +,\ +$ &$-$&$+$&$+$ &$+$&$-$&$-$ &$+$&$-$&$-$ &$-$&$+$&$+$  \\  \hline
$-,\ +,\ -$ &$-$&$+$&$+$ &$+$&$-$&$-$ &$-$&$+$&$+$ &$+$&$-$&$-$   \\ \hline
$-,\ -,\ +$ &$-$&$+$&$+$ &$-$&$+$&$+$ &$+$&$-$&$-$ &$+$&$-$&$-$   \\ \hline
$-,\ -,\ -$ &$-$&$+$&$+$ &$-$&$+$&$+$ &$-$&$+$&$+$ &$-$&$+$&$+$   \\ \hline
$+,\ +,\ +$ &$+$&$+$&$+$ &$+$&$+$&$+$ &$+$&$+$&$+$ &$+$&$+$&$+$   \\ \hline
$+,\ +,\ -$ &$+$&$+$&$+$ &$+$&$+$&$+$ &$-$&$-$&$-$ &$-$&$-$&$-$   \\ \hline
$+,\ -,\ +$ &$+$&$+$&$+$ &$-$&$-$&$-$ &$+$&$+$&$+$ &$-$&$-$&$-$   \\ \hline
$+,\ -,\ -$ &$+$&$+$&$+$ &$-$&$-$&$-$ &$-$&$-$&$-$ &$+$&$+$&$+$   \\ \hline
%\\ \hline
\end{tabular}
\caption[t1]{Signs of the Killing form eigenvalues in 
function of the signs of the parameters $\s,\,\LA,\,\la$.}
\label{tableA1}
\end{table}
%%%%%%%%%%%%%%%%%%%%%%%%%%%%%%%%%%%%%%%%%%%%%%%%%
%%%%%%%%%%%%%%%%%%%%%%%%%%%%%%%%%%%%%%%%%%%%%%%%%%%%%%%%%%%%%%%%%%%%%%%%%%%%
\begin{table}[hbt]
\centering
\begin{tabular}{|c||c|c|}
\hline
 $\boldsymbol{\mbox{\bf Signs of }\s,\LA,{\la}}$ & \textbf{Compact subalgebras} & \textbf{Basis of generators}
 \\ \hline%\hline 
 $-,\ +,\ +$ & so(4) &$J^0,P^1,P^2;\,R^0,Q^1,Q^2$ \\  \hline
 $-,\ +,\ -$ & so(4) &$J^0,P^1,P^2;\,Q^0,R^1,R^2$ \\ \hline
 $-,\ -,\ +$ & so(4) &$J^0,Q^1,Q^2;\,P^0,R^1,R^2$ \\ \hline
 $-,\ -,\ -$ & u(1) $\oplus$ u(1) $\oplus$ u(1) $\oplus$ u(1)  &$J^I,P^I,Q^I,R^I;I=0\ {\rm or}\ 1\ {\rm or}\ 2$ \\ \hline
 $+,\ +,\ +$ & so(4) $\oplus$ so(4) & $J^I,P^I,Q^I,R^I;I=0,1,2$ \\ \hline
 $+,\ +,\ -$ & so(4) &$J^I,P^I;I=0,1,2$ \\ \hline
 $+,\ -,\ +$ & so(4) &$J^I,Q^I;I=0,1,2$ \\ \hline
 $+,\ -,\ -$ & so(4) &$J^I,R^I;I=0,1,2$ \\ \hline
%\\ \hline
\end{tabular}
\caption[t1]{Compact sub-algebras and their basis of generators in 
function of the signs of the parameters $\s,\,\LA,\,\la$.
}
\label{tableA2}
\end{table}
%%%%%%%%%%%%%%%%%%%%%%%%%%%%%%%%%%%%%%%%%%%%%%%%%
In Table~\ref{tableA2}, we only show one possibility of so(4) sub-algebra for each choice of signs. But, in the first line for instance, there is another so(4) sub-algebra spanned by $\{J^0,Q^1,Q^2;\,R^0,P^1,P^2\}$, with similar alternatives for the other case.
It may be useful to note that we have the following sub-algebras, so(3) or so(1,2) depending of the signs of  $\s,\,,\LA,\,\la$: 
\[\ba{llll}
\{J^0,\,P^1,\,P^2\}: &[J^0,P^1]=P^2\,,\ &[P^1,P^2]=\LA J^0\,,\ 
&[P^2,J^0]=P^1\,;\esp
\{J^0,\,Q^1,\,Q^2\}: &[J^0,Q^1]=Q^2\,,\ &[Q^1,Q^2]=\la J^0\,,\ 
&[Q^2,J^0]=Q^1\;\esp
\{J^0,\,R^1,\,R^2\}: &[J^0,R^1]=R^2\,,\ &[R^1,R^2]=\s\LA\la J^0\,,\ 
&[R^2,J^0]=R^1\,.
\ea\]

\noindent \textbf{Remark:} Going back to our example -- corresponding 
to the first line of each Table -- we remark  that we have a de 
Sitter sub-algebra so(1,3) spanned by $\bsJ,\bsP$, with positive 
cosmological constant $\LA>0$, another one spanned by 
 $\bsJ,\bsQ$, with positive cosmological constant ${\la}>0$, and also 
an anti-de Sitter\footnote{This sub-algebra possesses an own sub-algebra 
so(3) spanned by $J^0$, $P^1$, $P^1$, which shows that it is really de 
Sitter so(1,3), and not anti-de Sitter so(2,2).} 
 sub-algebra so(2,2) spanned by $\bsJ,\bsR$, with negative cosmological 
constant $\s\LA {\la}<0$ . 
A similar remark applies to the other choices for the signs of the 
parameters $\s,\LA,\la$.

%%%%%%%%%%%%%%%%%%%%%%%%%%%%%%%%%%%%%%%%%%%%%%%%%%%%%%%%%%%%%%
\subsubsection{Factorization}\label{factorization}
We check now that, in the three cases displayed in the first 
three lines of Tables \ref{tableA1} and \ref{tableA2}, the 
\sads\ algebra factorizes in two de Sitter algebras so(1,3).   
To see this explicitly, let us consider first the case of the first line
of these tables,
and define two triplets of generators
\[
(X^i,\,i=1,2,3) \equiv ({J'}^0,{P'}^1,{P'}^2)\,,\quad
(Y^i,\,i=1,2,3) \equiv (-{P'}^0,{J'}^1,{J'}^2)\,,
\]
where we use the normalized \sads\ generators
\[
{J'}^I=J^I\,,\quad{P'}^I=P^I/\sqrt{|\LA|}\,,\quad
{Q'}^I=Q^I/\sqrt{|\la|}\,,\quad{R'}^I=R^I/\sqrt{|\LA\la|}\,.
\]
The $X$'s and $Y$'s obey the canonical so(1,3) commutation relations
\[
[X^i,\,X^j] = \e^{ijk} X^k\,,\quad [X^i,\,Y^j] = \e^{ijk} Y^k
\,,\quad [Y^i,\,Y^j] = -\e^{ijk} X^k\,.
\]
Thus $(J^I,\,I=0,1,2;\,P^I,\,I=0,1,2)$ $\equiv$
$(T_\a,\,\a=1,\cdots,6)$
 form just another basis for
the same algebra so(1,3), with commutation relations which one may
write as
\[
[T_\a,T_\b] = f_{\a\b}{}^\g T_\g\,.
\] 
Now it is a matter of checking that
$(-{R'}^I;\,{Q'}^I$ $\equiv$
$(S_\a,\,\a=1,\cdots,6)$ obey together with the $T$'s the commutation
relations
\[
[T_\a,S_\b] = f_{\a\b}{}^\g S_\g\,\,\quad [S_\a,S_\b] = f_{\a\b}{}^\g T_\g\,.
\]
The first ones show that the $S$'s span the adjoint representation of
so(3,1), and the second ones express the closure of the algebra \sads\
generated by the 12 generators $T_\a$, $S_\a$. This set of commutation
rules is of the type shown in \equ{SGG-algebra}, with $C=1$. $C$ being
positive, the factorization \equ{SGG-factorized-alg} holds: the \sads\
algebra splits in two   de Sitter  factors:
\eq 
\sads = \ds_+ \oplus \ds_- = {\rm so}(1,3)_+ \oplus {\rm so}(1,3)_-\,,
\eqn{A-factorization}
generated by
\eq 
J^I_\pm = \half \lp J^I \mp \frac{R^I}{\sqrt{\LA \la}} \rp \,,\quad
P^I_\pm = \half \lp \frac{P^I}{\sqrt{\LA}} \pm \frac{Q^I}{\sqrt{{\la}}}\,, \rp
\eqn{generators-ds-pm}
with the commutation rules
\[
[J^I_\pm, J^J_\pm] = \e^{IJ}{}_K J^K_\pm\,,\quad
[J^I_\pm, P^J_\pm] = \e^{IJ}{}_K P^K_\pm\,,\quad
[P^I_\pm, P^J_\pm] = -\e^{IJ}{}_K J^K_\pm\,.
\]
The cases corresponding to the second or third lines of Tables
\equ{tableA1} and \equ{tableA2} are equivalent and can be deduced from
the first case by interchanging the $Q$'s with the $R$'s, or 
the $P$'s with the $R$'s, respectively. The Riemannian cases ($\s=1$) 
displayed in 
the three last lines  of the tables follow equivalent patterns.
The explicit results are
summarized in Table \equ{tableA3}.
 %%%%%%%%%%%%%%%%%%%%%%%%%%%%%%%%%%%%%%%%%%%%%%%%%%%%%%%%%%%%%%%%%%%%%%%%%%%
\begin{table}[hbt]
\centering
\begin{tabular}{|c|c|ll|}
\hline
 {\bf Signs of } & && \\ 
$\boldsymbol{\s,\LA,\la}$ &{\bf Factorization} &{\bf Generators}& \\ 
\hline%\hline 
$-,\ +,\ +$ &${\rm so}(1,3)_+ \oplus {\rm so}(1,3)_-$&$J^I_\pm = \half \lp J^I \mp \frac{R^I}{\sqrt{\LA \la}} \rp$ ,
                                                     &$P^I_\pm = \half \lp \frac{P^I}{\sqrt{\LA}} \pm \frac{Q^I}{\sqrt{{\la}}} \rp$  \\  \hline
$-,\ +,\ -$ &${\rm so}(1,3)_+ \oplus {\rm so}(1,3)_-$&$J^I_\pm = \half \lp J^I \pm \frac{Q^I}{\sqrt{-\la}} \rp$ ,
                                                     &$P^I_\pm = \half \lp \frac{P^I}{\sqrt{\LA}} \pm \frac{R^I}{\sqrt{{-\LA\la}}} \rp$  \\  \hline
$-,\ -,\ +$ &${\rm so}(1,3)_+ \oplus {\rm so}(1,3)_-$&$J^I_\pm = \half \lp J^I \pm \frac{P^I}{\sqrt{-\LA}} \rp$ , 
                                                     &$P^I_\pm = \half \lp \frac{Q^I}{\sqrt{ \la}} \pm \frac{R^I}{\sqrt{{-\LA\la}}} \rp$  \\  \hline
$-,\ -,\ -$ &${\rm so}(2,2)_+ \oplus {\rm so}(2,2)_-$&$J^I_\pm = \half \lp J^I \pm \frac{P^I}{\sqrt{-\LA}} \rp$ , 
                                                     &$P^I_\pm = \half \lp \frac{Q^I}{\sqrt{-\la}} \pm \frac{R^I}{\sqrt{{ \LA\la}}} \rp$
\\ \hline
$+,\ +,\ +$ &${\rm so}(4)_+ \oplus {\rm so}(4)_-$    &$J^I_\pm = \half \lp J^I \pm \frac{P^I}{\sqrt{\LA}} \rp$ , 
                                                     &$P^I_\pm = \half \lp \frac{Q^I}{\sqrt{\la}} \pm \frac{R^I}{\sqrt{{\LA\la}}} \rp$
\\  \hline
$+,\ +,\ -$ &${\rm so}(1,3)_+ \oplus {\rm so}(1,3)_-$&$J^I_\pm = \half \lp J^I \pm \frac{P^I}{\sqrt{\LA}} \rp$ , 
                                                     &$P^I_\pm = \half \lp \frac{Q^I}{\sqrt{-\la}} \pm \frac{R^I}{\sqrt{{-\LA\la}}} \rp$  \\  \hline
$+,\ -,\ +$ &${\rm so}(1,3)_+ \oplus {\rm so}(1,3)_-$&$J^I_\pm = \half \lp J^I \pm \frac{Q^I}{\sqrt{\la}} \rp$ , 
                                                     &$P^I_\pm = \half \lp \frac{P^I}{\sqrt{-\LA}} \pm \frac{R^I}{\sqrt{{-\LA\la}}} \rp$  \\  \hline
$+,\ -,\ -$ &${\rm so}(1,3)_+ \oplus {\rm so}(1,3)_-$&$J^I_\pm = \half \lp J^I \mp \frac{R^I}{\sqrt{\LA\la}} \rp$ , 
                                                     &$P^I_\pm = \half \lp \frac{P^I}{\sqrt{-\LA}} \pm \frac{Q^I}{\sqrt{{-\la}}} \rp$  \\  \hline
\end{tabular}
\caption[t1]{Factorization properties of $\sads$ in 
function of the signs of the parameters $\s,\,\LA,\,\la$.}
\label{tableA3}
\end{table}

%%%%%%%%%%%%%%%%%%%%%%%%%%%%%%%%%%%%%%%%%%%%%%%%%%%%%%
\subsubsection{Invariant quadratic forms}

The \sads\ algebra has four quadratic Casimir operators
$C_i = C_i^{\a\b}\TT_\a\TT_\b\,,\ i=1,2,3,4$:
\[\ba{c}
C_1 = J^I J_I + \dfrac{P^I P_I}{\s\LA} + \dfrac{Q^I Q_I}{\s {\la}}
+ \dfrac{R^I R_I}{\LA \la}\esp
C_2 = J^I P_I + \dfrac{Q^I R_I}{\s \la}\,,\quad 
C_3 = J^I Q_I + \dfrac{P^I R_I}{\s\LA} \,,\quad 
C_4 = J^I R_I + P^I Q_I\,,
\ea\]
to which correspond four invariant quadratic forms $K^i_{ab}$ 
proportional to the inverse matrices $(C^{-1})^i_{ab}$ (we only write their non-vanishing components):
\eq\ba{l}
K^1_{J^I,J^J} = \eta_{IJ}\,,\quad K^1_{P^I,P^J} = \s\LA\eta_{IJ}\,,\quad 
 K^1_{Q^I,Q^J} = \s {\la}\eta_{IJ}\,,\quad 
    K^1_{R^I,R^J} = \LA {\la}\eta_{IJ}\,,\esp
K^2_{J^I,P^J} = \eta_{IJ}\,,\quad K^2_{Q^I,R^J} = \s {\la}\eta_{IJ}\,,\esp
K^3_{J^I,Q^J} = \eta_{IJ}\,,\quad K^3_{P^I,R^J} = \s \LA\eta_{IJ}\,,\esp
K^4_{J^I,R^J} = \eta_{IJ}\,,\quad K^4_{P^I,Q^J} = \eta_{IJ}\,,
\ea\eqn{4-quad-forms}
all -- the fourth one excepted -- being non-degenerate only if both $\LA$ and $\la$ are 
non-vanishing. We note that the first one is the Killing form \equ{Killing}

All this is a generalization of the case of the \ads\ algebra, which has two invariant 
quadratic form~\cite{Witten:1988hc}. If we choose the \ads\ basis
$J^I$ and $P^I$, which obeys the commutation rules of the first line of 
\equ{SADS-algebra}, the two quadratic forms read:
\eq\ba{l}
k^1_{J^I,J^J} = \eta_{IJ}\,,\quad k^1_{P^I,P^J} = \s\LA\eta_{IJ}\,,\esp
k^2_{J^I,P^J} = \eta_{IJ}\,,\esp
\ea\eqn{2-quad-forms}
the first one being the Killing form of \ads, non-degenerate if $\LA\not=0$.

%%%%%%%%%%%%%%%%%%%%%%%%%%%%%%%%%%%%%%%%%%%%
\noindent{\bf Acknowledgments:} This work was partially funded by the Conselho Nacional de Desenvolvimento Cient\'{\i}fico e
   Tecnol\'{o}gico -- CNPq (Brazil).

%\end{document}
%%%%%%%%%%%%%%%%%%%%%%%%%%%%%%%%%%%%%%%%%%%%%%%%%%%%%%%%%%%%

%%%%%%%%%%%%%%%%%%%%%%%%%%%%%%%%%%%%%%%%%%%%%%%%%%%%%%%%%%%%%

\small
\begin{thebibliography}{99}

\bibitem{BCOP} R.M.S. Barbosa, C.P. Constantinidis, Z. Oporto and O. Piguet,
``Quantization of Lorentzian 3d Gravity by Partial Gauge Fixing'',
\cqg{29}{155011}{2012}, [arXiv: 1204.5455 [gr-qc]].

\bibitem{barbero-immirzi} J.F. Barbero,  ``Reality conditions and Ashtekar
variables: A Different perspective'', Phys. Rev. D51 (1995) 5507, 
[arXiv: gr-qc/9410013]; \\
     Giorgio Immirzi, ``Real and complex connections for canonical
gravity'',\\ Class.Quant.Grav. 14 (1997) L177-L181, 
[arXiv: gr-qc/9612030]. 

\bibitem{holst} S. Holst, ``Barbero's Hamiltonian derived from a generalized 
Hilbert-Palatini action'', \pr{D53}{1996}{5966}, 
[arXiv: gr-qc/9511026].

\bibitem{bonzom-livine} V. Bonzom and E.R. Livine, 
``A Immirzi-like parameter for 3d quantum gravity''
\cqg{25}{2008}{195024}, [arXiv:0801.4241[gr-qc]]

\bibitem{Witten:1988hc}
Edward Witten.
{``(2+1)-Dimensional Gravity as an Exactly Soluble System''},
{\em Nucl. Phys.}, B311:46, 1988.

\bibitem{Geiller1} Marc Geiller, Karim Noui, 
``Testing the imposition of the Spin Foam Simplicity Constraints'', 
 \cqg{29}{2012}{135008}, [arXiv:1112.1965 [gr-qc]].
\bibitem{Geiller2} Marc Geiller, Karim Noui, 
``A note on the Holst action, the time gauge, and the Barbero-Immirzi parameter'', \GRG{45}{2013}{1733}, [arXiv:1212.5064 [gr-qc]].
\bibitem{Geiller3} Jibril Ben Achour, Karim Noui, Chao Yu, 
``Testing the role of the Barbero-Immirzi parameter and the choice of connection in Loop Quantum Gravity, [arXiv:1306.3241 [gr-qc]].

\bibitem{Buffenoir} E. Buffenoir, K. Noui and Ph. Roche, ``Hamiltonian quantization of Chern-Simons theory with SL(2,$\complex$) group'',
\cqg{19}{2002}{4953}, [arXiv: hep-th /02/02121].

\bibitem{Carlip} S.Carlip, ``Quantum Gravity in 2+1 dimensions'', Cambridge Monographs on Mathematical Physics (1998).

\bibitem{carlip-gegenberg} S. Carlip and J. Gegenberg, 
``Gravitating topological matter in 2+1 dimensions'', \pr{D44}{424}{1991}.

\bibitem{carlip-gegenberg-mann} S. Carlip, J. Gegenberg and R.B. Mann, 
``Black holes in three-dimensional topological gravity'',
\pr{D51}{6854}{1995}.

\bibitem{freidel-mann-popescu} L. Freidel, R.B. Mann and E.M. Popescu, 
``Canonical analysis of the BCEA topological matter model coupled to gravitation in (2+1) dimensions'', \cqg{22}{3363}{2005}.

\bibitem{general-ref} C. Rovelli, ``Quantum Gravity'', Cambridge Monography on Math.
	Physics  (2004);\\
A. Ashtekar and J. Lewandowski, ``Background independent quantum gravity: 
        A status report'', \cqg{21}{2004}{R53}) [arXiv:gr-qc/0404018];\\
T. Thiemann,
 	``Modern Canonical Quantum General Relativity'',
   Cambridge Monographs on Mathematical Physics (2008);\\
M. Han, W. Huang and Y. Ma ``Fundamental structure of loop quantum  gravity'',
   \ijmp{D16}{2007}{1397}, [arXiv:gr-qc/0509064].

\bibitem{dirac}   P.A.M. Dirac, ``Lectures on Quantum Mechanics'',
Dover, 2001;   \\
M. Henneaux, C. Teitelboim, ``Quantization of Gauge Systems'',
Princeton University Press, 1994.

\bibitem{Constantinidis-Luchini-Piguet} C.P. Constantinidis, G. Luchini and O. Piguet,
``The Hilbert space of Chern-Simons theory on the cylinder.
A Loop Quantum Gravity approach'', \cqg{27}{2010}{065009},
[arXiv:0907.3240[gr-qc]].

\bibitem{quantiz-of-CS} G.V. Dunne, R. Jackiw, C.A. Trugenberger,
  ``Chern-Simons Theory in the Schr\"o\-din\-ger Representation'',
  \annp{194}{1989}{197}; \\
E.Guadagnini, M.Martellini, M.Mintchev, ``Braids and Quantum Group Symmetry in 
Chern-Simons Theory'', \np{B336}{1990}{581};\\
Steven Carlip, ``Quantum Gravity in 2+1 Dimensions'', 
Cambridge Monographs on Mathematical Physics (2003).

\bibitem{witten-WZW}  E. Witten, ``Nonabelian Bosonization in 
Two Dimensions'',\\ \cmp{92}{984}{455}.


\end{thebibliography}
\end{document}